\documentclass{article}
\usepackage[left=1in,right=1in,top=1in,bottom=1in]{geometry}
\usepackage{authblk}
\usepackage{graphicx,xcolor}
\usepackage{ulem,url}
\usepackage[title]{appendix}
\usepackage{cite}
\usepackage{amsmath}
\DeclareMathOperator*{\argminB}{argmin}

\usepackage{mathtools}  
\usepackage{xfrac}
\usepackage{amsthm,amssymb,bm}
\numberwithin{equation}{section}

\title{An epidemical model with nonlocal spatial infections}
\author[1]{Su Yang}
\author[1]{Weiqi Chu}
\author[1]{Panayotis G. Kevrekidis}
\affil[1]{Department of Mathematics and Statistics, University of Massachusetts, Amherst, 01003-4515, Massachusetts, USA}
\date{}

\begin{document}

\maketitle

\begin{abstract}
The SIR model is one of the most prototypical compartmental models in epidemiology. Generalizing this ordinary differential equation (ODE) framework into a spatially distributed partial differential equation (PDE) model is a considerable challenge. In the present work, we extend a recently proposed model based on nearest-neighbor spatial interactions by one of the authors in~\cite{vaziry2022modelling} towards a nonlocal, nonlinear PDE variant of the SIR prototype. We then seek to develop a set of tools that provide insights for this PDE framework. Stationary states and their stability analysis offer a perspective on the early spatial growth of the infection. Evolutionary computational dynamics enable visualization of the spatio-temporal progression of infection and recovery, allowing for an appreciation of the effect of varying parameters of the nonlocal kernel, such as, e.g., its width parameter. These features are explored in both one- and two-dimensional settings. At a model-reduction level, we develop a sequence of interpretable moment-based diagnostics to observe how these reflect the total number of infections, the epidemic's epicenter, and its spread. Finally, we propose a data-driven methodology based on the sparse identification of nonlinear dynamics (SINDy) to identify approximate closed-form dynamical equations for such quantities. These approaches may pave the way for further spatio-temporal studies, enabling the quantification of epidemics.
\end{abstract}

\section{Introduction}
The COVID-19 pandemic has resulted in staggering numbers of infections and fatalities over the past four years. According to the World Health Organization (WHO) dashboard, the number of reported cases worldwide exceeds 775 million, with over 7 million deaths. In the United States alone, there have been 103 million cases and over 1.2 million fatalities. Such devastating numbers underscore the need for significantly advanced preparedness for future pandemics to monitor and accurately predict disease spread and implement effective intervention strategies. A major role in managing the pandemic was played by the rapid development of mRNA-based vaccines, which saved many lives despite the emergence of newer and sometimes more aggressive COVID variants, such as the delta and omicron variants in late 2020 and late 2021, respectively.

From a mathematical perspective, the modeling of infections through compartmental models has a rich history dating back a century to the seminal work of \cite{kermack}. Over time, these developments have been extensively documented in reviews and books, such as \cite{hethcote,castillo2011,chen2015analyzing}. The advent of the COVID-19 pandemic has intensified efforts within the applied mathematics community and beyond to develop predictive tools for understanding the temporal evolution and associated risks of pandemics. Notably, comparative studies across modeling efforts, like \cite{ForeCast}, have provided valuable perspectives. Several reviews have already summarized diverse modeling approaches~\cite{cao2022covid,shakeel}, focusing on aspects such as network metapopulation models~\cite{review_meta}. Despite significant progress, as various works~\cite{bertozzi2020,holmdahl2020} have indicated, numerous challenges remain unaddressed and that---as the corresponding Lancet article~\cite{lance} indicated---the relevant pandemic is ``far from over''.

In addition to widely used metapopulation models (see, e.g., \cite{vespi2008,arenas2020,review_meta}), researchers have developed spatio-temporal resolution models to study the progression of epidemics \cite{kevrekidis2021, mammeri2020, viguerie2021}. Some of these models employ reaction-diffusion approaches \cite{kevrekidis2021, mammeri2020}, incorporating possibly time-dependent diffusivities to simulate various mitigation measures that restrict social interactions. Others emphasize the role of inflows from neighboring regions \cite{viguerie2021} and introduce the concept of population mobility \cite{jiang2024mobility}. Related approaches integrating vaccination dynamics can be found in studies such as \cite{pathogens12010088}. Recently, one of the authors developed a spatio-temporal model from the first principles of agent dynamics (and subsequent coarse graining), incorporating agent mobility and interactions that transmit infection among nearby agents \cite{vaziry2022modelling}. This approach yielded a nonlinear diffusion PDE model at the continuum limit, representing a generalization of the standard well-mixed SIR ODE case.

Our objective in this study is to extend the framework proposed by Vaziry et al. \cite{vaziry2022modelling}. Instead of restricting individuals to nearby locations only, we account for transportation to more distant areas using a decaying kernel that reflects typical commuting preferences. This approach leads to the development of an integro-differential, nonlocal variant of the SIR model, which potentially captures the spatio-temporal dynamics more comprehensively. We explore this model in both 1+1 and 2+1 spatio-temporal dimensions, providing a detailed description in section \ref{sec: model}, followed by an analysis of stationary states and their stability in section \ref{sec: stability} to understand early epidemic emergence and spread dynamics. Section \ref{sec: evolution} focuses on systematic visualizations of space-time dynamics for susceptible, infected, and recovered populations, highlighting infection waves under varying kernel parameters such as the kernel width. In section \ref{sec: moment}, we employ a moment-based approach to analyze system dynamics, offering insights into infection epicenters and their spread, driven by empirical data. Finally, we summarize our findings and outline future research directions in section \ref{sec: summary}.

\section{Model description} \label{sec: model}

\subsection{Nonlocal infectious spread}
Classic SIR models typically assume a homogeneous population in space (``well-mixed assumption''), ignoring the impact of spatial relations on disease spread. However, many epidemics show directional tendencies and significant spatial dependence~\cite{vespi2008,arenas2020,badr2020association,glaeser2022jue,rapti2023role}. It is essential to understand how epidemics unfold in nonhomogeneous populations and how this spatial variation influences the overall spread of disease. Here, we adopt a modeling-minded approach to develop relevant distributed SIR variants and explore the corresponding phenomenology numerically.
Our aim is to analyze a suite of modeling and computation tools at the level of partial and ordinary differential equations for a recently proposed model of spatio-temporal infection evolution~\cite{vaziry2022modelling}. A natural next step involves integrating these models with spatial data to enhance their applicability.

Let $S(x,t)$, $I(x,t)$, and $R(x,t)$ be the population densities at location $x\in\mathbb{R}^d$ ($d=1$ or $2$, typically) and time $t$ for the susceptible ($S$), infected ($I$), and recovered ($R$) individuals, respectively. We assume that the total population is time-independent, denoted by $f(x)$, and satisfies
\begin{equation}
    f(x) = I(x,t) + S(x,t) + R(x,t), \quad \int_{\mathbb{R}^d} f(x)~\mathrm{d}x = 1.
    \label{eq: mass_conservation}
\end{equation}
In addition, the three densities are governed by the generalized
integro-differential variant of the dynamical equations of~\cite{vaziry2022modelling} according to:
\begin{equation}\label{eq: nonlocal_SIR} 
    \begin{aligned}
    S_t(x,t) &= -\beta S(x,t)\int_{\mathbb{R}^d} a(x,y)I(y,t)~\mathrm{d}y , \\
    I_t(x,t) &= \beta S(x,t)\int_{\mathbb{R}^d} a(x,y)I(y,t)~\mathrm{d}y  -\gamma I(x,t), \\
    R_t(x,t) &= \gamma I(x,t),
    \end{aligned}
\end{equation}
where $\beta$ is the infection rate, $\gamma$ is the recovery rate, and $a(x,y)$ is a spatial infection kernel that represents the probability density that an infected individual at location $y$ infects a susceptible individual at location $x$. 
In the classic SIR model, each infected individual infects $\beta$ individuals per unit of time. Therefore, we impose that for any value of $y$, the integral over $\mathbb{R}^d$ of $a(x,y)$ with respect to $x$ equals 1, i.e., we normalize
the relevant kernel:
\begin{equation}
\int_{\mathbb{R}^d} a(x,y)~\mathrm{d}x = 1,
\end{equation}
ensuring that an infected individual still infects $\beta$ individuals, albeit distributed non-homogeneously across space.
We let the spatial interaction take the form of a translation-invariant kernel
\begin{equation} \label{eq: kernel}
a(x,y) = \phi(|x-y|),
\end{equation}
where the kernel $\phi(|r|)$ decays as the distance $|r|$ increases. This implies that the infection is solely determined by the distance between two individuals. 
We assume that populations in close proximity exhibit a stronger tendency for infection, implying direct mobility in ``physical space'' between regions. While we make this assumption for simplicity in this initial exposition, we recognize that the concept of ``effective distance''~\cite{brock} (influenced by various forms of transportation) may be relevant for future studies.

\subsection{Connections to the spatially local SIR model}
By using the translation-invariant kernel~\eqref{eq: kernel}, we establish a fundamental connection between the spatially nonlocal SIR models \eqref{eq: nonlocal_SIR} and the local models derived in \cite{vaziry2022modelling}, which represent the continuum limits of nearest-neighbor interactions. We use the change of variables $\xi = y - x$ and obtain
\begin{equation}
\begin{aligned}
       \int_{\mathbb{R}^d}\phi(|x-y|)I(y,t)~\mathrm{d}y 
    = I(x,t) + \frac12 I_{xx}(x,t)\int_{\mathbb{R}^d}\xi^2\phi(|\xi|) ~\mathrm{d}\xi+\int_{\mathbb{R}^d}\phi(|\xi|) O(\xi^4)~\mathrm{d}\xi.
\end{aligned}
\end{equation}
When the function $\phi(|r|)$ is predominantly concentrated around $r=0$, the nonlocal model approximates the spatially local SIR model in \cite{vaziry2022modelling} (after we drop the high-order small terms)
\begin{equation}\label{eq: local_SIR}
    \begin{aligned}
    S_{t}(x,t) &= -\beta S(x,t)I(x,t) - \mu S(x,t)I_{xx}(x,t), \\
    I_{t}(x,t) &= \beta S(x,t)I(x,t) + \mu S(x,t)I_{xx}(x,t) -\gamma I(x,t), \\
    R_{t}(x,t) &= \gamma I(x,t), 
    \end{aligned}
\end{equation}
where $\mu=\beta\sigma^2/2$ and $\sigma^2$ is the second-order moment of the density $\phi$. 
This formulation demonstrates how, for small $\sigma$, the nonlocal model simplifies to a local model with an added {\it nonlinear} diffusion term, capturing the spatial spread of infections.

It is important to note an observation that applies to both nonlocal and local cases. Specifically, the models examined here not only conserve the total population in an integral sense, as per Equation~\eqref{eq: mass_conservation}, but also conserve it locally at every $x$ (i.e., $S(x,t)+I(x,t)+R(x,t)$ is conserved 
for all spatial positions $x$ and does not depend on time $t$). This implies an assumption of ``shorter-term'' mobility, where individuals move around but return to their base, rather than a scenario involving migration from one location to another. In future considerations, and in line with the mobility patterns discussed in \cite{arenas2020,rapti2023role}, it might be relevant to incorporate time-dependent transport terms. These terms would conserve the total population globally but not necessarily locally. Having set up the model and clarified its main assumptions, we now analyze its mathematical features.

\section{Linear stability analysis}\label{sec: stability}
As $S(x,t)$, $I(x,t)$, and $R(x,t)$ are related by the conservation of population \eqref{eq: mass_conservation}, we proceed in the linear stability analysis with the governing equations of only two densities: the susceptible and infected population densities $S(x,t)$ and $I(x,t)$. We consider the linear stability around the equilibrium state
\begin{equation}
    I(x,t)=0, \quad S(x,t)=S_0(x),
    \label{uniform stationary solutions}
\end{equation}
where no infection exists. In the linear stability framework, $I(x,t)$ and $S(x,t)$ may grow exponentially with respect to time $t$ following a small initial perturbation of the epidemic. We formulate their initial growth using 
\begin{align} \label{eq: ansatz}
    S(x,t) &= S_0(x) + \varepsilon S_1(x)e^{\lambda t}, \quad I(x,t) = \varepsilon I_{1}(x) e^{\lambda t},
\end{align}
where $\lambda$ denotes the leading eigenvalue of the linearized system of \eqref{eq: nonlocal_SIR}, and $S_1(x)$ and $I_1(x)$ are the corresponding eigenfunctions associated with the leading eigenvalue $\lambda$, representing the spatial pattern of growth in $S(x,t)$ and $I(x,t)$ initially.

\subsection{Spatially local SIR models}
We substitute \eqref{eq: ansatz} into the local SIR model \eqref{eq: local_SIR} and keep the leading-order terms, obtaining
\begin{equation}\label{eq: local eigenvalue system}
     \begin{aligned}
     \mu S_{0}(x)I_{1}''(x) + \beta S_{0}(x)I_{1}(x) - \gamma I_1(x) &=  \lambda I_{1}(x),  
    \\
     -\beta S_{0}(x)I_{1}(x) - \mu S_{0}(x)I_{1}''(x) &= \lambda S_{1}(x).
    \end{aligned}
\end{equation}
The eigenvalue $\lambda$ signals potential instability when $\lambda > 0$, with the corresponding eigenfunction $I_1(x)$ indicating the spatial direction in which the infection grows. This growth, in turn, decreases the susceptible population along the spatial direction of $S_1(x)$. Given that the eigenfunction $I_1(x)$ reflects the spatial epidemic growth from a background vanishing value, we are particularly interested in nonnegative eigenfunctions $I_1(x)$ to ensure that $I(x,t)$ stays nonnegative as a population density. We refer to nonnegative eigenfunctions as epidemiologically relevant ones.

We study the eigenvalue problem \eqref{eq: local eigenvalue system} for both $d=1$ and $d=2$ with the boundary condition
\begin{equation}
\lim_{|x|\to\infty} I_1(x) = 0.
\end{equation}
In numerical simulations, we employ homogeneous Dirichlet boundary conditions for sufficiently large computational domains. Recall that $S_0(x)$ can be arbitrarily chosen. For the case $d=1$, we choose two representative susceptible population densities, which are a Gaussian density $S_0^\text{G}$ and a periodic density $S_0^\text{p}$:
\begin{equation} \label{eq: S0}
    \begin{aligned}
    S_0^\text{G}(x) &\propto \exp \left[-\left(x - 3\right)^{2}\right], \qquad
    S_0^\text{p}(x) \propto 3\sin(3x) + 3.1.
    \end{aligned}
\end{equation} 
Our choice reflects the expectation that localized blobs represent significant population concentrations for $S_0^\text{G}$, while a periodic decomposition is suited for more complex population profiles in the case of $S_0^\text{p}$ as the potential Fourier decomposition of these more intricate patterns. We normalize the density $S_0$ so that it initially integrates to 1 on the computational domain.

\begin{figure}[ht]
    \centering
    \includegraphics[width=0.9\linewidth]{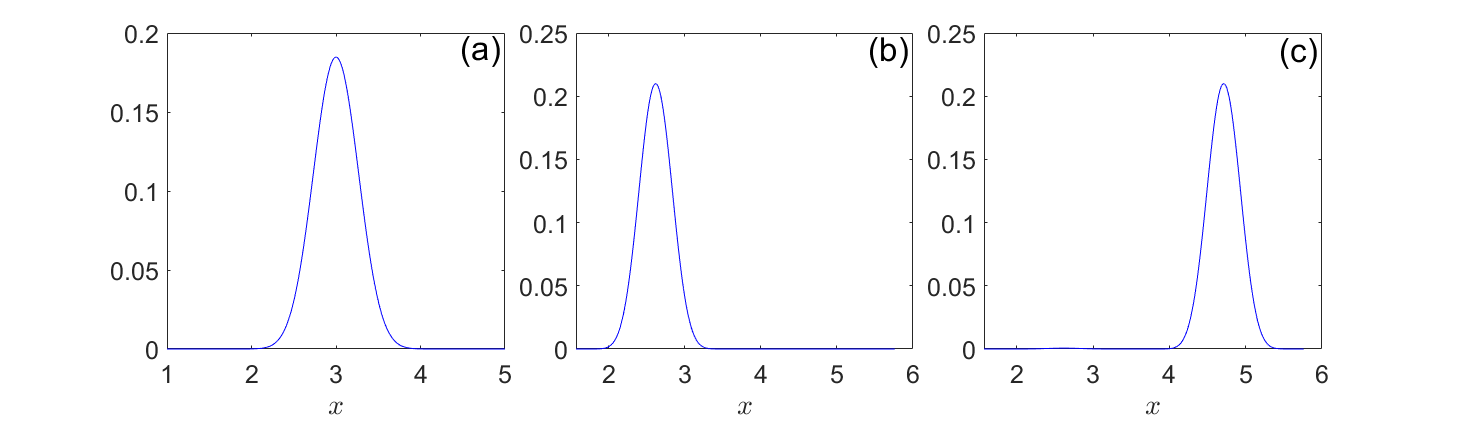}
    \caption{Eigenfunctions $I_1(x)$ of the local SIR models \eqref{eq: local_SIR} with (a) the Gaussian initial density $S_0^{\text{G}}$ and (b--c) the periodic density $S_0^{\text{p}}$. For all simulations, we use the same parameters $\beta = 0.8, \gamma = 0.1, \mu = \beta\sigma^{2}/2$, where $\sigma = 0.1$, and the computational domains for the two cases are $\left[1,5\right]$ and $\left[\pi/2,11\pi/6\right]$, respectively. 
    }
    \label{fig:1}
\end{figure}

We show the eigenfunctions $I_1(x)$ of the eigenvalue problem \eqref{eq: local eigenvalue system} in Figure \ref{fig:1}. For the Gaussian initial density $S_0^\text{G}$ in \eqref{eq: S0}, we only find one epidemiologically relevant eigenfunction, corresponding to the largest eigenvalue $\lambda = 0.322$. For the periodic initial density $S_0^\text{p}$ in \eqref{eq: S0}, we find two corresponding feasible eigenfunctions in the problem (and domain) under consideration, both associated with the largest eigenvalue $\lambda = 0.236$, due to the degeneracy associated with the two effective blobs in the initial density. The corresponding eigenvectors reflect the potential growth in either one or the other blob, while the remaining one subsides, as shown in Figure \ref{fig:1}.

\subsection{Spatially nonlocal SIR models}
\begin{figure}[htp]
    \centering
    \includegraphics[width=0.99\linewidth]{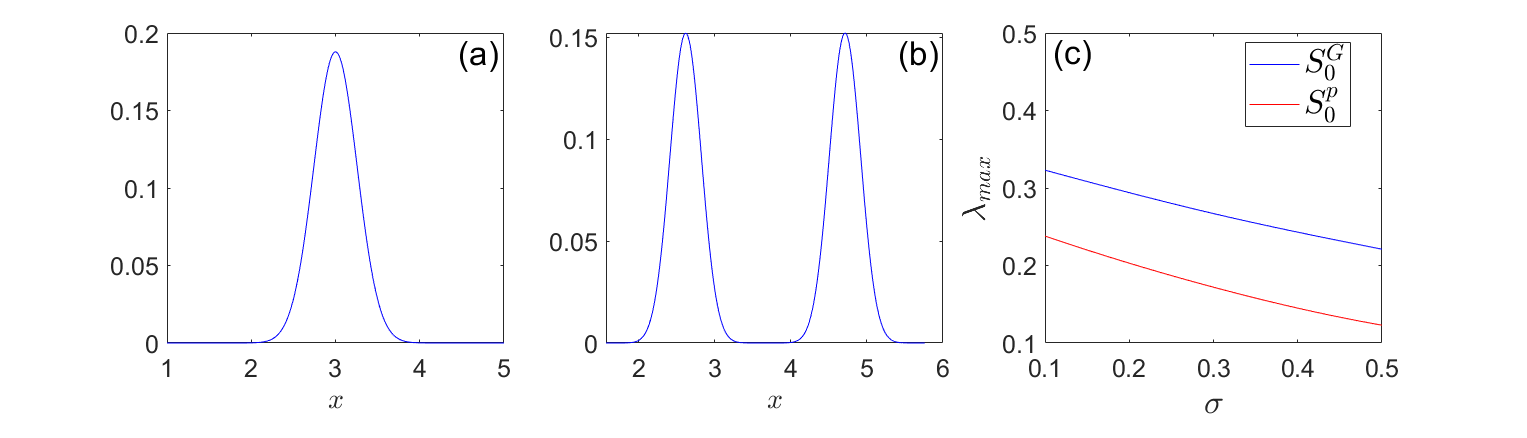}
    \caption{(a--b): Eigenfunctions of the nonlocal model with $\sigma=0.1$ and different initial densities in \eqref{eq: S0}. The corresponding eigenvalues are $\lambda = 0.323$ for (a) $S_0^{\text{G}}$ and $\lambda = 0.238$ for (b) $S_0^{\text{p}}$. (c): Leading eigenvalues varying with the kernel width parameter $\sigma$. In all simulations, we fix parameters $\beta = 0.8$ and $\gamma = 0.1$.}
    \label{fig:2}
\end{figure}

In a similar fashion as before, we substitute the spectral stability expansion of \eqref{eq: ansatz} into the nonlocal SIR model \eqref{eq: nonlocal_SIR} and obtain that
\begin{equation}\label{eq: nonlocal eigenvalue problem}
    \begin{aligned}
    -\gamma I_1(x) + \beta S_0(x)\int_{\mathbb{R}^d}a(x,y)I_{1}(y)~\mathrm{d}y  &= \lambda I_1(x), \\
    -\beta S_0(x)\int_{\mathbb{R}^d}a(x,y)I_{1}(y)~\mathrm{d}y  &= \lambda S_1(x).
    \end{aligned}
\end{equation}
We solve the above eigenvalue problem and show the eigenfunctions corresponding to the two densities $S_0$ in \eqref{eq: S0} in Figure \ref{fig:2}. We select the Gaussian kernel 
\begin{equation} \label{eq: 1d-kernel}
    a(x,y) = \phi(|x-y|) = \frac{1}{\sqrt{2\pi}\sigma} \exp\left( - \frac{|x-y|^2}{2\sigma^2}\right)
\end{equation}
with the kernel width parameter $\sigma=0.1$. 
For the Gaussian density $S_0^{\text{G}}$, we obtain one epidemiologically relevant eigenfunction, associated with the largest eigenvalue $\lambda = 0.323$. For the periodic density $S_0^{\text{p}}$, we only observe one epidemiologically relevant eigenfunction which is associated with the largest eigenvalue $\lambda = 0.238$. 
We then vary the kernel width parameter $\sigma$ and solve the eigenvalue problem \eqref{eq: nonlocal eigenvalue problem}. The leading eigenvalues are shown in Figure \ref{fig:2}(c). As $\sigma$ increases, the maximum eigenvalue decreases for both initial densities $S_0$ in \eqref{eq: S0}. This trend is linked to the normalization of our kernel: as the kernel becomes narrower, the infection probability is higher in the immediate vicinity, leading to a more rapidly developing spatial wave of infections. Conversely, with a larger $\sigma$, the infection probability is weaker nearby and stronger at greater distances, making it less likely for localized populations to generate a rapidly growing infection. We also observe that when the kernel width $\sigma$ is small, the nonlocal SIR model \eqref{eq: nonlocal_SIR} closely approximates the local SIR model \eqref{eq: local_SIR}, which results in similar leading eigenvalues in the linear stability analysis. 

For the case $d = 2$, the linear stability analysis is analogous to the $d = 1$ case, ultimately resulting in a more computationally expensive variant of the above eigenvalue problems. To offer a prototypical sense of the results, we choose parameters $\beta = 100$, $\gamma = 0.5$, and $\sigma = 0.1$ and consider the linear stability around two equilibrium densities, the Gaussian and periodic densities:
\begin{equation}
    S_0^{\text{G}}(x_1,x_2) \propto \exp\left[-\left(x_1 - 3\right)^{2} - \left(x_2 - 3\right)^{2}\right], \qquad S_0^{\text{p}}(x_1,x_2) \propto 3\sin(3x_1) + 3.1.
\end{equation}
When the equilibrium density is Gaussian ($S_0^{\text{G}}$) [in this case, centered at the point $(3,3)$], both the local and nonlocal SIR models have only one eigenvalue associated with an epidemiologically relevant eigenfunction. The eigenvalues are $27.160$ for the local model and $27.269$ for the nonlocal model. Similarly, when the equilibrium density is periodic ($S_0^{\text{p}}$), both models again have one single eigenvalue corresponding to an epidemiologically relevant eigenfunction, with eigenvalues of $\lambda = 9.520$ for the local model and $\lambda = 9.557$ for the nonlocal model. 
Notice that our periodic state is quasi-one-dimensional, as it does not depend on the variable $x_2$. In both settings, cases with sufficiently small $\sigma$ exhibit eigenvalues and growth rates similar to those of the local models. The significantly larger eigenvalues in the $d=2$ case are due to the higher infection parameter $\beta$ used to accelerate the spread of the epidemic in the numerical simulations. The decrease in the relevant eigenvalue with increasing $\sigma$ can be explained similarly to the previous paragraph. Linear stability analysis also reveals that in the $d=2$ case, a smaller kernel width parameter $\sigma$ leads to a closer alignment between the nonlocal model and the corresponding local model.

\section{Evolution dynamics}\label{sec: evolution}
We now turn to the detailed dynamical comparison of the nonlocal and local SIR models (\eqref{eq: nonlocal_SIR} and \eqref{eq: local_SIR}) for both $d=1$ and $d=2$ cases. We use the initial conditions:
\begin{equation}\label{eq: Initial Conditions}
    \begin{aligned}
    I(x,0) = \eta I_0(x), \quad  S(x,0) = (1-\eta) S_0(x), \quad R(x,0) = 0,
    \end{aligned}
\end{equation}
where $\eta$ is the ratio of the infected population initially, and $I_0$ and $S_0$ are the spatial distributions of the infected and susceptible populations, respectively. We require that both $I_0$ and $S_0$ integrate to 1 and they are not necessarily the same. 
In the spatial domain, we use the pseudo-spectral method for the $d=1$ local models and the central finite difference method for the $d=2$ local models.
We have confirmed that the particular choice of numerical method does not
affect in any way the nature of our conclusions.
For the nonlocal model, we apply the composite rectangle rule to approximate the relevant integral. We perform the time integration using the fourth-order Runge--Kutta method.

\subsection{One-dimensional spatial SIR models}
\begin{figure}[ht]
    \centering
    \includegraphics[width=0.85\linewidth]{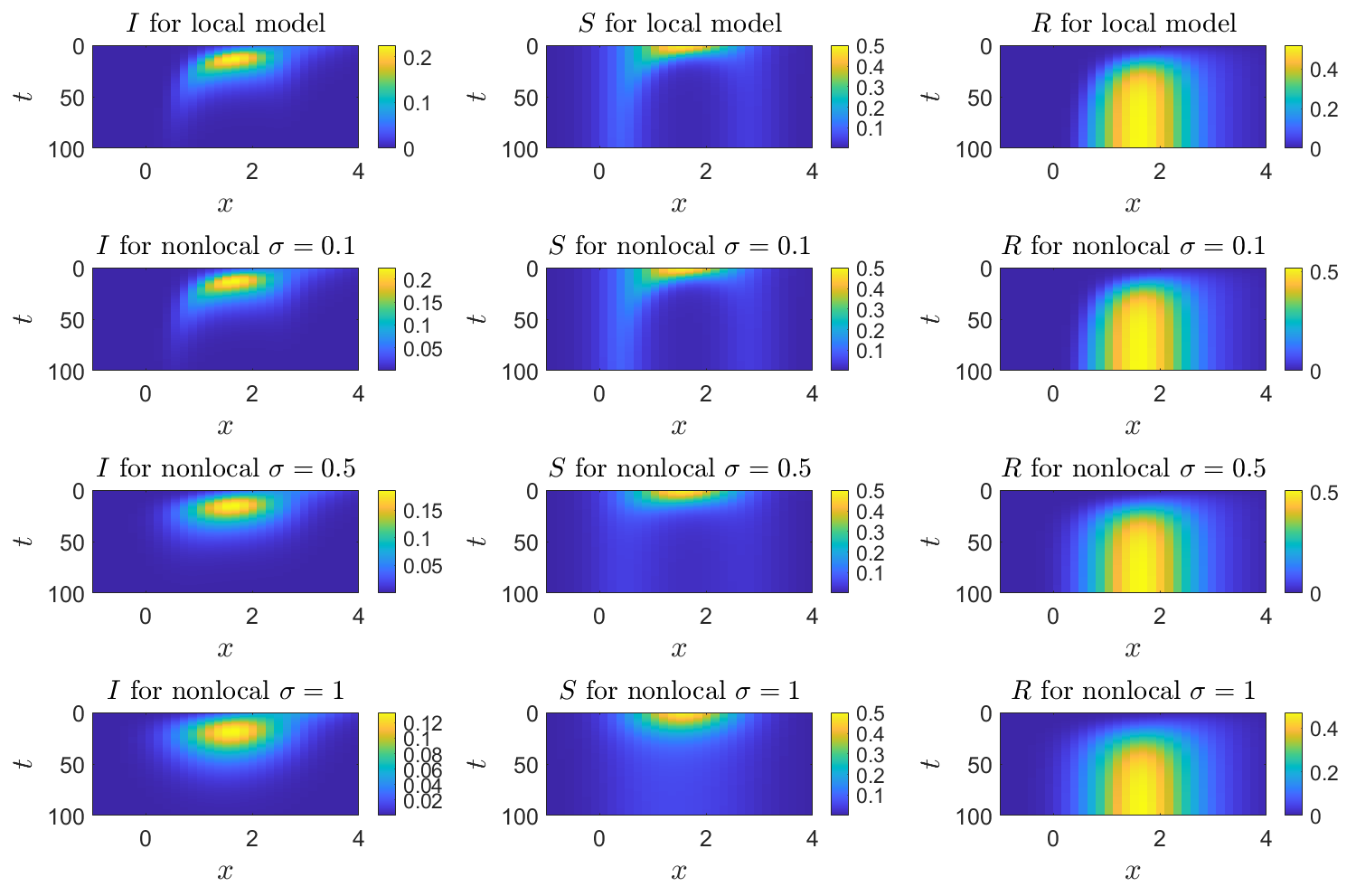}
    \caption{Evolution dynamics of the spatially local SIR model with $\sigma=0.1$ (the top panel) and spatial nonlocal SIR models with different width parameters $\sigma=0.1$, $0.5$, $1$ (the bottom three panels). We fix the initial infection ratio $\eta=0.1$, infection rate $\beta=0.8$, recovery rate $\gamma=0.1$, and $\mu = \beta\sigma^{2}/2$. 
    }    
    \label{fig: 1d-SIR}
\end{figure}

We investigate the epidemic evolution for both the local and nonlocal SIR models. We use the Gaussian initial conditions:
\begin{equation} \label{eq: 1d_Cs}
    \begin{aligned}
    I_0(x) & \propto \exp\left[-\left(x-2.5\right)^{2}\right], \quad S_0(x) \propto \exp\left[-\left(x-1.5\right)^{2}\right],
    \end{aligned}
\end{equation}
where we deliberately let the infected and susceptible populations centered at different locations. We vary the kernel parameter $\sigma$ in \eqref{eq: 1d-kernel} and examine how nonlocality influences the epidemic spreading. We show the evolution dynamics of both local and nonlocal models in Figure \ref{fig: 1d-SIR}. 

In this example, we select the epicenter of the infection at a location significantly separated from the area with the highest population density. Initially, a small group of infected individuals is clustered around $x=2.5$, while the susceptible population is centered at $x=1.5$. As the dynamics unfold, the infection spreads to the left, creating a densely infected region around $x=1.5$, where most of the susceptible population was originally concentrated. Over time, the number of infected individuals decreases as the disease subsides, and those who were infected transition into the recovered status. 
At each spatial location, we observe infections starting from 0, peaking, and then declining, following similar dynamics to the classical non-spatial SIR models. The local SIR model \eqref{eq: local_SIR} closely approximates the nonlocal SIR model \eqref{eq: nonlocal_SIR} when the kernel width $\sigma$ is small.
In both the local case (and similarly in the weakly nonlocal case), it is interesting to note that there exists a ``wave of infection'' that outskirts to locations of $x < 1.5$. This coincides with the areas containing the largest susceptible populations over extended periods. We will return to this point when we explore the moments of the system. Ultimately, in all scenarios, the population transitions to a uniform distribution of recovered individuals.

As the kernel width $\sigma$ increases, infections reach the far-left region earlier compared to the local models or nonlocal models with smaller kernel widths. This is because nonlocal models facilitate long-range infections. Normalizing the interaction kernel ensures that each infected individual affects the same number of susceptible individuals across different models. In the nonlocal models, infected individuals are distributed more widely across space, whereas in the local models, they are more concentrated.
In the case where $\sigma=1$, we also observe that infections emerge more slowly near the peak of the susceptible population (around the $x=1.5$ region) compared to cases with smaller kernel widths.
In the local model or nonlocal models with smaller kernel widths $\sigma$, the dynamics of the susceptible population near $x=1.5$ exhibit a vacuous region and show pronounced spatial heterogeneity at $t > 25$. This occurs because when infections reach the population center around $x=1.5$, infected individuals become contagious and further propagate the disease to nearby individuals in the vicinity. This reinforcement mechanism concentrates the infection near the population center, leading to rapid infection of all susceptible individuals within a short period.
In contrast, for $\sigma=1$, the susceptible population displays a smooth spatial transition for $t > 25$ and a more uniform distribution across space. This is attributed to infections being spread more broadly, resulting in smoother spatial distributions.
Similar discussions are provided in Appendix \ref{sec: 1d-periodic} for the case where the centers of the initial densities $S_0$ and $I_0$ are co-located.

\subsection{Two-dimensional spatial SIR models}
\begin{figure}[htp]
    \centering
    \includegraphics[width=0.55\linewidth]{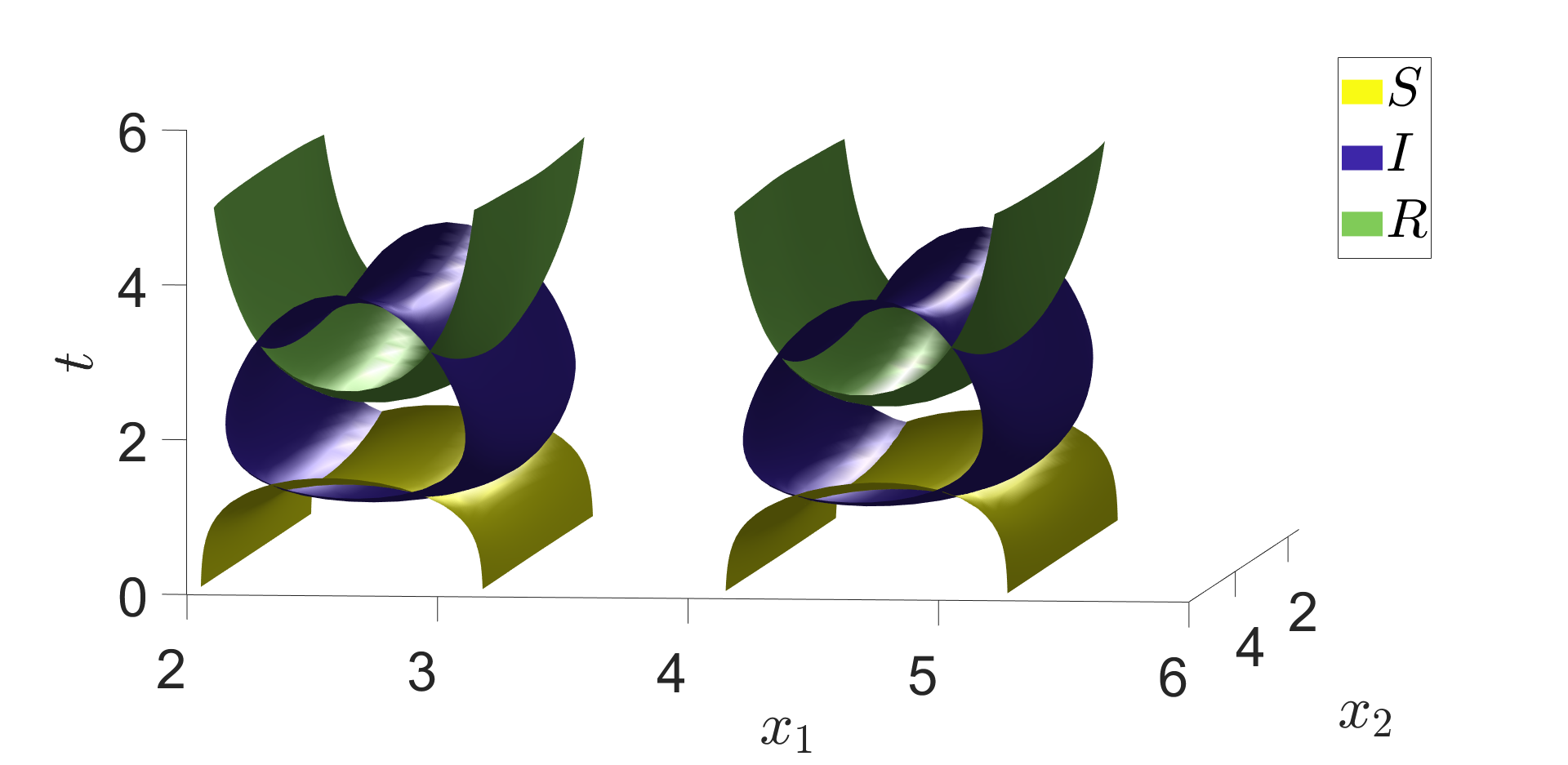}
    \caption{The iso-surface representation of the evolution dynamics of the 2d nonlocal SIR model \eqref{eq: nonlocal_SIR} with the width parameter $\sigma = 1$. The infection rate is $\beta = 100$ and the recovery rate $\gamma = 0.5$. The initial infection ratio $\eta = 0.01$. 
    The initial conditions $S_0(x)$ and $I_0(x)$ are given in \eqref{eq: 2d-initial} and the computational domain is $\left[\pi/2, 11\pi/6\right]\times\left[\pi/2, 11\pi/6\right]$. The three population densities ($S(x,t)$, $I(x,t)$, and $R(x,t)$) have constant function values $0.05, 0.035$, and $0.045$ on their respective iso-surfaces. 
    }
    \label{fig:4}
\end{figure}

We now consider the $d=2$ case, using periodic initial conditions for the susceptibles and a localized ``blob'' of infection:
\begin{equation}\label{eq: 2d-initial}
    \begin{aligned}
        S_0(x_1,x_2) \propto 3\sin\left(3x_1\right) + 3.1, \quad I_0(x_1,x_2) \propto \exp\left[-\left(x_1-\frac{7\pi}{6}\right)^{2} - \left(x_2- \frac{7\pi}{6}\right)^{2}\right].
    \end{aligned}
\end{equation}
The susceptible population is periodic in the $x_1$-direction and homogeneous in the $x_2$-direction, forming two quasi-one-dimensional blobs centered at $x_1 = 5\pi/6$ and $x_1 = 3\pi/2$. The infected population is initially centered at $(7\pi/6, 7\pi/6)$, situated in the middle of the two blobs of the susceptible population, leading to symmetric infections in the two blobs. We plot the iso-surfaces of the evolution dynamics of the 2d nonlocal SIR model with $\sigma=1$ in Figure \ref{fig:4}. 

In the iso-surface of the susceptible population with an iso-value of $0.050$, we observe that the susceptible population decreases and stabilizes as the majority becomes infected around $t=1.5$. Since the infection source is initially located at $x_2 = 7\pi/6$, the $x_2$ distribution of the populations is centered around $x_2 = 7\pi/6$, indicating a faster reduction in the susceptible population at this location compared to nearby regions. On the other hand, the $x_1$ epicenter of the infection dynamics is clearly at the location of the blobs, i.e., at $x_1 = 5\pi/6$ and $x_1 = 3\pi/2$. 
The infected iso-surface corresponds to the value $I(x,t)=0.035$. The infected population begins to appear (in terms of its iso-contours) before the susceptible population subsides around $t = 1.25$, and it eventually disappears around $t = 4$, at which time the recoveries have taken over;
the latter two have also originated from $(5\pi/6,7\pi/6)$ and $(3\pi/2,7\pi/6)$, as is clearly discerned in Figure~\ref{fig:4}. Initially, the infected population appears also near $(5\pi/6, 7\pi/6)$ and $(3\pi/2, 7\pi/6)$, with the $x_1$-coordinates aligned with the initial centers of the susceptible blobs and the $x_2$-coordinate with the initial infection center. The water-drop shape of the iso-surface illustrates two distinct phases of the infected population: growth and decline, representing the infection and recovery processes, respectively. Notably, the infection growth rate in the first phase is faster than the recovery rate in the second phase. 
The recovered iso-surface corresponds to a constant value of $R(x,t) = 0.045$. The recovered population begins to appear at $t = 2.5$ as the infected population decays in the initial infection spots. It continues to grow in the surrounding regions until it encompasses the entire susceptible population.

In numerical simulations, we observe that a smaller kernel width parameter $\sigma$ results in behavior closer to that of the local model. For the local model (or the nonlocal models with small $\sigma$ values) with the same initial conditions \eqref{eq: 2d-initial}, the iso-surfaces resemble those in Figure \ref{fig:4}. However, the concave surfaces of the susceptible iso-surfaces are more pronounced in the local model due to stronger local infection effects. We also explore a 2d weakly nonlocal spatial model with Gaussian initial conditions and present the iso-surface plots in Appendix \ref{sec: 1d-periodic}.

\subsection{Decay speed of the susceptible population}
\begin{figure}[htp]
    \centering
    \includegraphics[width=0.7\linewidth]{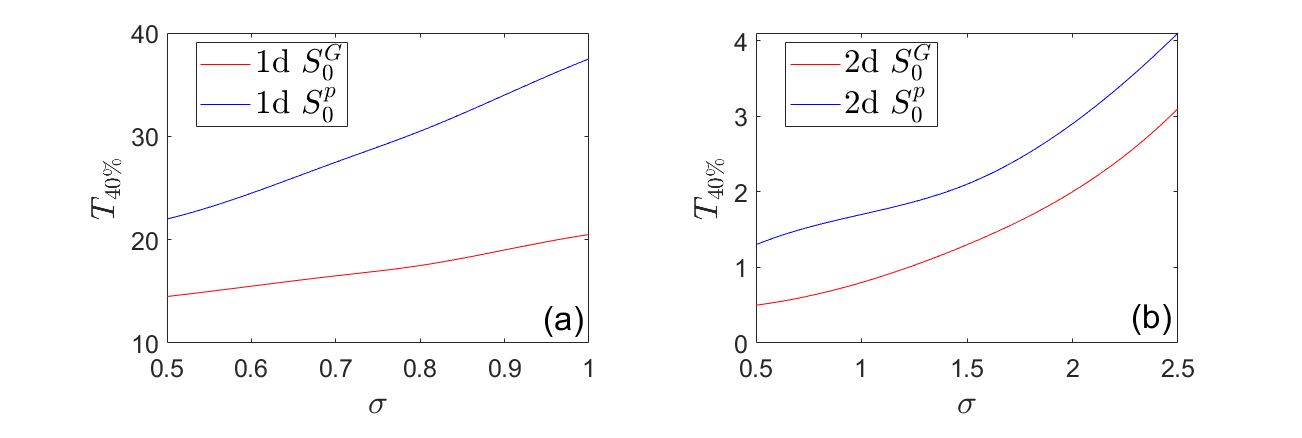}
    \caption{The decay time $T_{40\%}$ changing with the width parameter $\sigma$ for: (a) 1d and (b) 2d nonlocal SIR models \eqref{eq: nonlocal_SIR}. Among all simulations, we fix the initial conditions and all parameters except for $\sigma$.}
    \label{fig: T10}
\end{figure}

In this section, we examine how nonlocality affects the rate at which the susceptible population decreases. As a concrete diagnostic, we examine the infinity norm of the susceptible population density in space (i.e., $\|S(\cdot,t)\|_{\infty}$) and record the time that $\|S(\cdot,t)\|_{\infty}$ decays to 40\% of its initial value, which we denote by $T_{40\%}$. While this percentage choice is not particularly special, it is representative of the time scales of the decay dynamics due to the infection. We fix other parameters and show how $T_{40\%}$ changes with the kernel width parameter $\sigma$ in \eqref{eq: 1d-kernel} in Figure \ref{fig: T10}.

For the 1d and 2d nonlocal SIR models, we observe that $T_{40\%}$ increases as the kernel width parameter $\sigma$ increases, indicating that it takes longer for the susceptible population to decline with larger widths. Specifically, the local model is most efficient in transmitting the infection, as intuitively expected, and as was previously explained.
In the local model \eqref{eq: local_SIR}, infections spread quickly to areas of high population density, where infected individuals affect nearby susceptible individuals, leading to rapid decay of the susceptible population in these dense regions. In contrast, in the nonlocal models, infected individuals infect susceptible individuals more evenly across space, thereby delaying the depletion of the susceptible population, particularly in localized scenarios considered here.
For progressively larger width parameters $\sigma$, infections spread more widely and more evenly, resulting in a later time for the maximum susceptible density to decay.


\section{Moment dynamics and their data-driven inference}\label{sec: moment}
Moment dynamics offer insights into the overall behavior and distribution of infected and susceptible populations over time. They are a valuable diagnostic tool for understanding epidemic progression in both temporal and spatial contexts, particularly in identifying patterns and rates of change. In this section, we investigate how the kernel width parameter $\sigma$ affects these dynamics in terms of moments, with the goal of inferring the underlying dynamics using a data-driven approach.
We define moments associated with a nonnegative function $g$ as
\begin{equation}\label{eq: moment quantities}
    \begin{aligned}
    Q_0^{g}(t) &= \int_{\mathbb{R}^d}g(x,t)~\mathrm{d}x,\qquad Q_k^{g}(t) = \frac{\int_{\mathbb{R}^d}x^kg(x,t)~\mathrm{d}x}{Q_0^{g}(t)}, ~~ k=1,2,\ldots,
    \end{aligned}
\end{equation}
where $g$ is replaced by one of the three population density functions ($S$, $I$, or $R$) in this section. The zeroth-order moment $Q_0^{g}$ is the total mass of population $g$, $Q_1^{g}$ is its center of population, and $Q_2^{g}$ relates to the population density's variance. More concretely, the variance $V(t)$ is given by
\begin{eqnarray}
    V(t)=\int_{\mathbb{R}^d} \frac{(x-Q_1^{g})^2 g(x,t)}{Q_0^{g}}~\mathrm{d}x= Q_2^{g}-(Q_1^{g})^2.
    \label{var}
\end{eqnarray}

\subsection{Evolution dynamics of moments}
\begin{figure}[htp]
    \centering
    \includegraphics[width=0.8\linewidth]{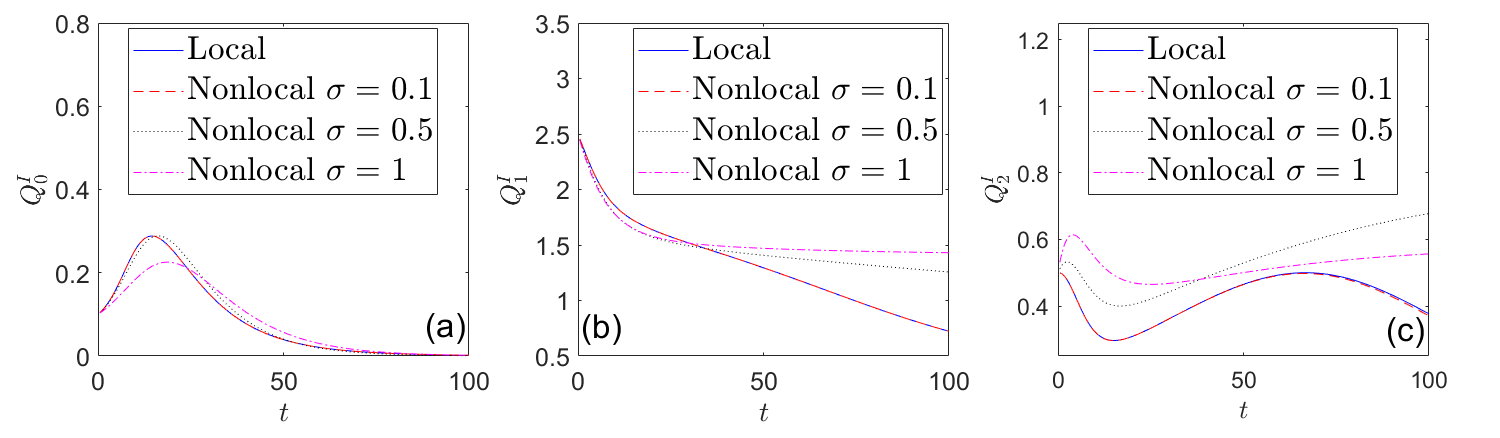} \\
    \includegraphics[width=0.8\linewidth]{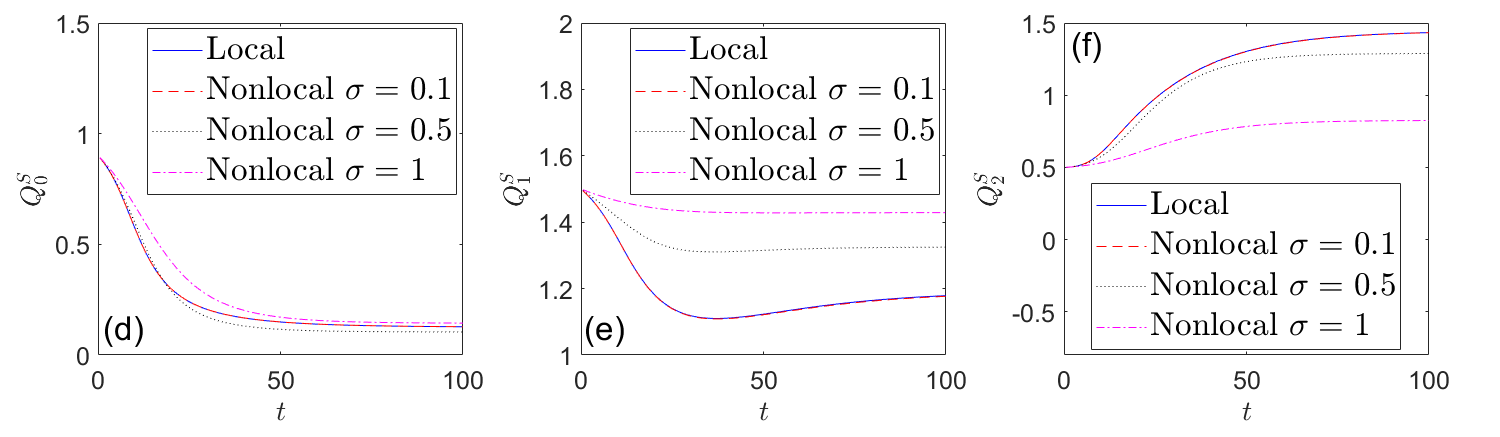}   
    \caption{Moment dynamics of different models for (a)--(c) the infected population and (d)--(f) the susceptible population. For all simulations, we fix the parameters $\beta = 0.8$ and $\gamma = 0.1$. We use $\mu = \beta\sigma^{2}/2$ and $\sigma = 0.1$ for the local model and use $\sigma = 0.1$, $0.5$, $1$ for the nonlocal model. 
    }
    \label{fig: moments}
\end{figure}

We solve the 1d nonlocal \eqref{eq: nonlocal_SIR} and local \eqref{eq: local_SIR} SIR models using the Gaussian initial conditions in \eqref{eq: 1d_Cs}, and present the dynamics of the moments associated with the infected and susceptible populations in Figure \ref{fig: moments}. Notably, the moment dynamics of the local model align closely with that of the nonlocal model when the kernel width $\sigma$ is set to a small value of $0.1$.

We observe that the total infected population, $Q_0^I$, initially rises as the epidemic spreads but then declines towards zero as most individuals transition from infection to recovery, consistent with the classic SIR models. In the local model or the nonlocal models with small kernel widths, the infection predominantly affects nearby individuals more intensely compared to the more evenly distributed effect in the nonlocal models with larger kernel widths. Consequently, the infected population peaks earlier, around $t = 18$, in the local model and the nonlocal models with small $\sigma$, in contrast to the later peak observed in the nonlocal models with larger kernel widths. Conversely, in progressively more nonlocal scenarios (e.g. models with smaller $\sigma$ values), the overall infection intensity is weaker earlier on but persists for a longer duration.

We also observe that the center of mass of the infected population, $Q_1^I$, decreases over time. Initially centered at $x=2.5$, the infected population shifts leftward towards the center of the susceptible population near $x=1.5$ as the epidemic progresses. This movement slows down once the infection reaches the central region at $x=1.5$. While this saturation-like behavior is seen in the local models, consistent with the spatio-temporal evolution discussed earlier in Figure~\ref{fig: 1d-SIR}, there appears to be a residual wave of infection spreading towards smaller $x$ values. The moment dynamics, both for the local model and the nonlocal case with $\sigma=0.1$, clearly illustrate this trend.

In the nonlocal models, the variance of the infected population initially increases for a short period due to nonlocal spatial infections. Shortly afterward, the infections accumulate near the center of the susceptible population near $x=1.5$, reducing the second moment, over the period of high infectivity. As the first wave of infection passes, the infected individuals recover, and the number of infected individuals decreases, causing the variance to increase again.

For the susceptible population, its zeroth moment decreases as a result of the infection. The center of mass stabilizes after an initial decrease, influenced by the reduction of susceptibles at larger values of $x$ due to the higher concentration of infections in that spatial region. Ultimately, as the infection diminishes the susceptible population, this contributes to the growth of its second moment.

\subsection{Data-driven inference of moment dynamics}
The (local and) nonlocal SIR models presented herein describe the spread of epidemics through spatially distributed populations. Deriving closed-form dynamics for the moments of these population distributions is desirable but challenging. Each moment depends on higher-order moments, resulting in an infinite hierarchy of coupled equations that cannot be simplified without suitable closure approximations. The nonlinearities and spatial dependencies in the model further complicate these relationships.

To address this paucity of results regarding the (nonetheless well-defined as shown above) moment dynamics, we utilize a data-driven approach whereby we feed the time series of such moment quantities to a widely used package for the sparse identification of nonlinear dynamics (SINDy)~\cite{Kutz_SINDy}. SINDy constitutes a data-driven methodology that aims to identify governing equations from time-series data. More specifically, it constructs a feature matrix from a library of candidate functions and identifies the underlying dynamics by representing the system as a sparse combination of candidate functions. In particular, we represent the dynamics using
\begin{equation}
    \dot {\bm y} = \Theta({\bm y})\Xi,
    \label{Sparse representation of the ODE system}
\end{equation}
where $\Theta({\bm y})$ is a feature matrix formed by a library of candidate functions and $\Xi$ is a sparse coefficient vector. We obtain $\Xi$ by minimizing a loss function
\begin{equation}
    \Xi = \argminB_{\Xi'}\|\dot {\bm y} - \Theta({\bm y})\Xi'\|_2 + \lambda \|\Xi'\|_1.
    \label{Sparse optimization}
\end{equation}

\begin{figure}[htp] \label{fig: sindy}
    \centering
    \includegraphics[width=0.8\linewidth]{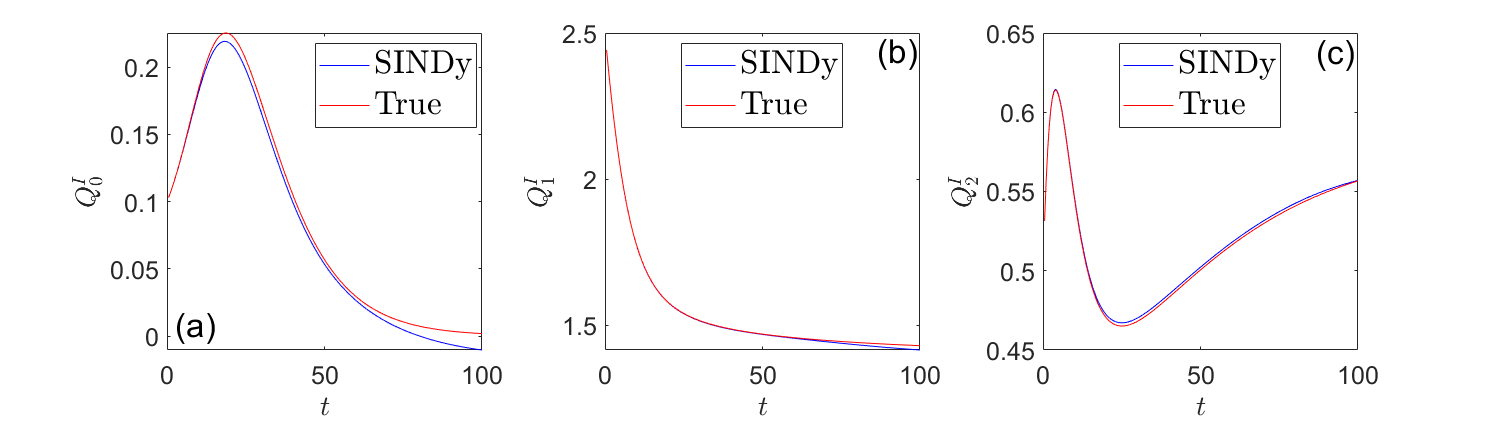}
    \includegraphics[width=0.8\linewidth]{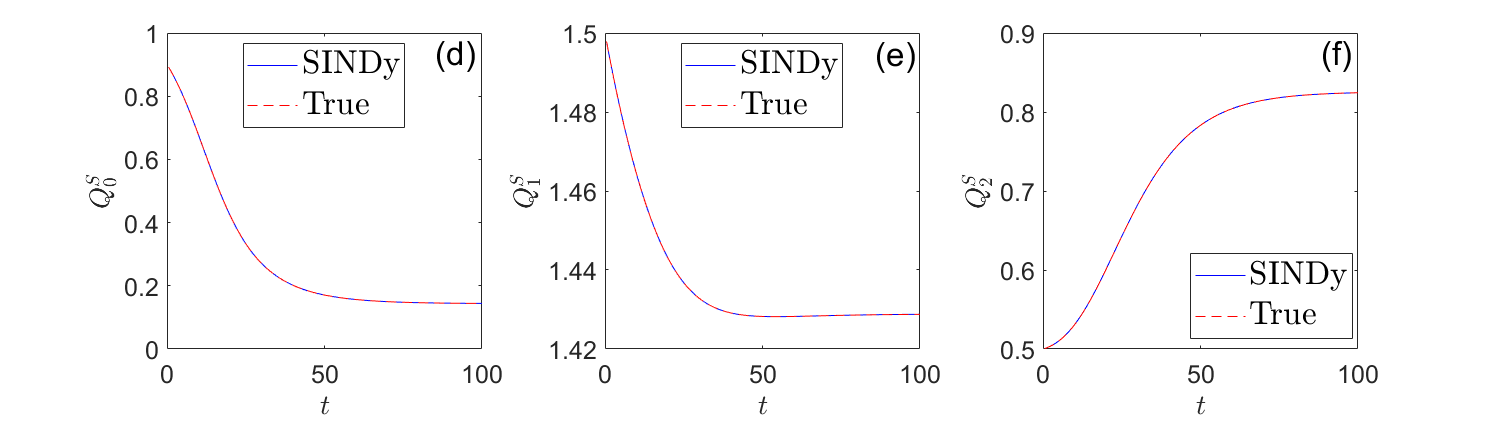}
     \caption{The true and learned moment dynamics of the infected and susceptible population: Figures (a)-(c) compare the SINDy learned moments with the true moments of the infected population, while (d)-(f) showcase the situation for the moments of the susceptible population.}
\end{figure}

We apply SINDy to learn the dynamics of the moments of the infected population, in which ${\bm y} = (Q_0^I, Q_1^I, Q_2^I)$. Given the nonlinear nature of the original models, we consider a library of three polynomials that includes linear, quadratic, and product terms (i.e., $a$, $a^2$, and $ab$). We solve the nonlocal model with $\sigma = 1$ for $I(x,t)$ and compute the corresponding moments to obtain the time series of ${\bm y}$. Using the time series of moments up to $t = 50$ as data, we learn the governing equations of the infected moments. Predictions are made for $t > 50$, and the predicted dynamics are compared with the true dynamics in Figure \ref{fig: sindy}(a--c).
The learned dynamics closely match the true moment dynamics of the infected population, both during the training period ($(0, 50]$) and the testing period ($50, 100]$). The inferred terms and coefficients of the inferred dynamics are detailed in Table \ref{tab: coefficients_I}.
\begin{table}[htp]
    \centering
    \begin{tabular}{c|c|c|c|c|c|c|c|c|c}
        &   $Q_0^I$   &  $Q_1^I$   &  $Q_2^I$   &  
        $\left(Q_0^I\right)^2$   &  $\left(Q_1^I\right)^2$    &  $\left(Q_2^I\right)^2$  &  
        $Q_0^IQ_1^I$  &  $Q_0^IQ_2^I$ &  $Q_1^IQ_2^I$ \\ \hline\hline 
    $\dot{Q}_0^I$ &  $-1.177$   &  $0.161$& $-0.294$& $-0.047$& $-0.065$& $0.091$& $0.903$& $-0.589$& $0.045$\\
    $\dot{Q}_1^I$ &    $0.093$&  $0.085$& $0.024$& $0$& $-0.046$& $-0.063$& $-0.004$& $-0.231$&$-0.029$ \\
    $\dot{Q}_2^I$ &$0.707$    &$-0.063$  &$0.160$ & $-0.290$& $0.027$& $-0.283$& $0.043$&$-1.637$ &$0.042$ \\
    \end{tabular}
    \caption{Coefficients and terms in the inferred moment dynamics of the infected population.}
    \label{tab: coefficients_I}
\end{table}

Similarly, we use SINDy to learn the dynamics of the susceptible moments, where ${\bm y}=(Q_0^S,Q_1^S,Q_2^S)$ in \eqref{Sparse optimization}. We use the true moment dynamics during the interval $(0, 60]$ as training data and then predict the dynamics for the interval $(60, 100]$. Figure \ref{fig: sindy}(d--f) shows the learned dynamics alongside the true dynamics. The learned dynamics closely overlap with the true dynamics, indicating that SINDy effectively infers the dynamics of the susceptible moments. Notice that we have found this to be the case
when considering the dynamics past the maximum of the epidemic inline, e.g., with the arguments of~\cite{Sauer21}. The inferred terms and coefficients of the inferred dynamics are detailed in Table \ref{tab: coefficients_S}.
\begin{table}[htp]
    \centering
    \begin{tabular}{c|c|c|c|c|c|c|c|c|c}
        &   $Q_0^S$   &  $Q_1^S$   &  $Q_2^S$   &  
        $\left(Q_0^S\right)^2$   &  $\left(Q_1^S\right)^2$    &  $\left(Q_2^S\right)^2$  &  
        $Q_0^SQ_1^S$  &  $Q_0^SQ_2^S$ &  $Q_1^SQ_2^S$ \\ \hline\hline 
    $\dot{Q}_0^S$ &  $3.133$   &  $-4.549$& $8.208$& $0.196$& $3.680$& $0.573$& $-1.881$& $-2.443$& $-6.748$\\
    $\dot{Q}_1^S$ &    $-0.014$&  $0$& $0.135$& $0.001$& $-0.005$& $-0.009$& $0.006$& $0.015$&$-0.081$ \\
    $\dot{Q}_2^S$ &$-0.660$    &$0.920$  &$-1.491$ & $0.001$& $-0.695$& $-0.143$& $0.362$&$0.389$ &$1.193$ \\
    \end{tabular}
    \caption{Coefficients and terms in the inferred moment dynamics of the susceptible population.}
    \label{tab: coefficients_S}
\end{table}

\section{Conclusions and future challenges}
\label{sec: summary}
In this work, we proposed a class of nonlocal analogs of the SIR models in space-time, developed as a generalization of the recent proposal of~\cite{vaziry2022modelling}. These models, in the limit of small variance, closely match the local models of the above work but also consider distributed interactions that decay with distance, generalizing the nearest-neighbor scenario originally considered. We performed linear stability analysis and obtained the dynamical evolution of such models for different degrees of nonlocality in both one- and two-spatial dimensions. We developed suitable visualizations of interest in their own right, particularly in two-dimensional settings, as they present spatial extensions of the well-known temporal epidemic compartmental evolution curves.

More importantly, we developed a series of diagnostics, including the different moments of the compartmental distributions and timescales associated with them, which enable us to perform a systematic comparison of the local and nonlocal spatial dynamics of the system and to intuitively explain the simulation observations. As a significant step in diagnostics, we set up a prototypical example of data-driven approaches to develop effective ODE models describing the epidemic moments. The successful realization of this approach offers promise for deriving relevant closed-form dynamical equations that describe, at a reduced level, the spatio-temporal evolution of epidemic dynamics.

Naturally, this effort suggests several directions for future study. 
Arguably the most important challenge is connecting the mathematical and computational analysis presented here with realistic data. While a kernel dependent on distance only may be a meaningful first approximation, it is also realistic to expect an anisotropic, directionally dependent kernel and to mathematically explore the quantitative impact of such dependencies. With suitable spatial distribution data of infections, it is crucial to identify (i.e., reverse engineer) the kernel that accurately describes spatial infectious interactions. Such an inference problem is explored in the setting of a mobility-based SIR model~\cite{jiang2024mobility}.

More broadly, developing and validating spatio-temporal models of epidemics at both smaller and larger scales--such as across multiple provinces, countries, and beyond--represents a crucial and intriguing direction for further epidemiological exploration. It is also conceivable that with appropriately distributed spatio-temporal data, machine learning methods such as PINNs~\cite{PINN}, deepXDE~\cite{xde_lorenz}, or similar approaches could be used to extract PDE-level models for spatial epidemic propagation. Such studies are currently underway and will be detailed in future publications.

\begin{appendices} 

\section{Evolution dynamics of the spatial SIR models}
\label{sec: 1d-periodic}

\subsection{A 1d example with co-localized initial densities}
In this section, we show more of the numerical simulations of the spatial SIR models for both 1d and 2d cases. We examine the evolution dynamics of a 1d example with the co-localized initial densities: 
\begin{equation} \label{eq: colocalization}
I_0(x) = S_0(x) \propto \exp\left[-\left(x-2.5\right)^{2}\right].
\end{equation}
The initial densities are different from the previous case in Figure \ref{fig: 1d-SIR}, where the centers of the two initial densities are significantly separated from each other. 
\begin{figure}[htp]
    \centering
    \includegraphics[width=0.85\linewidth]{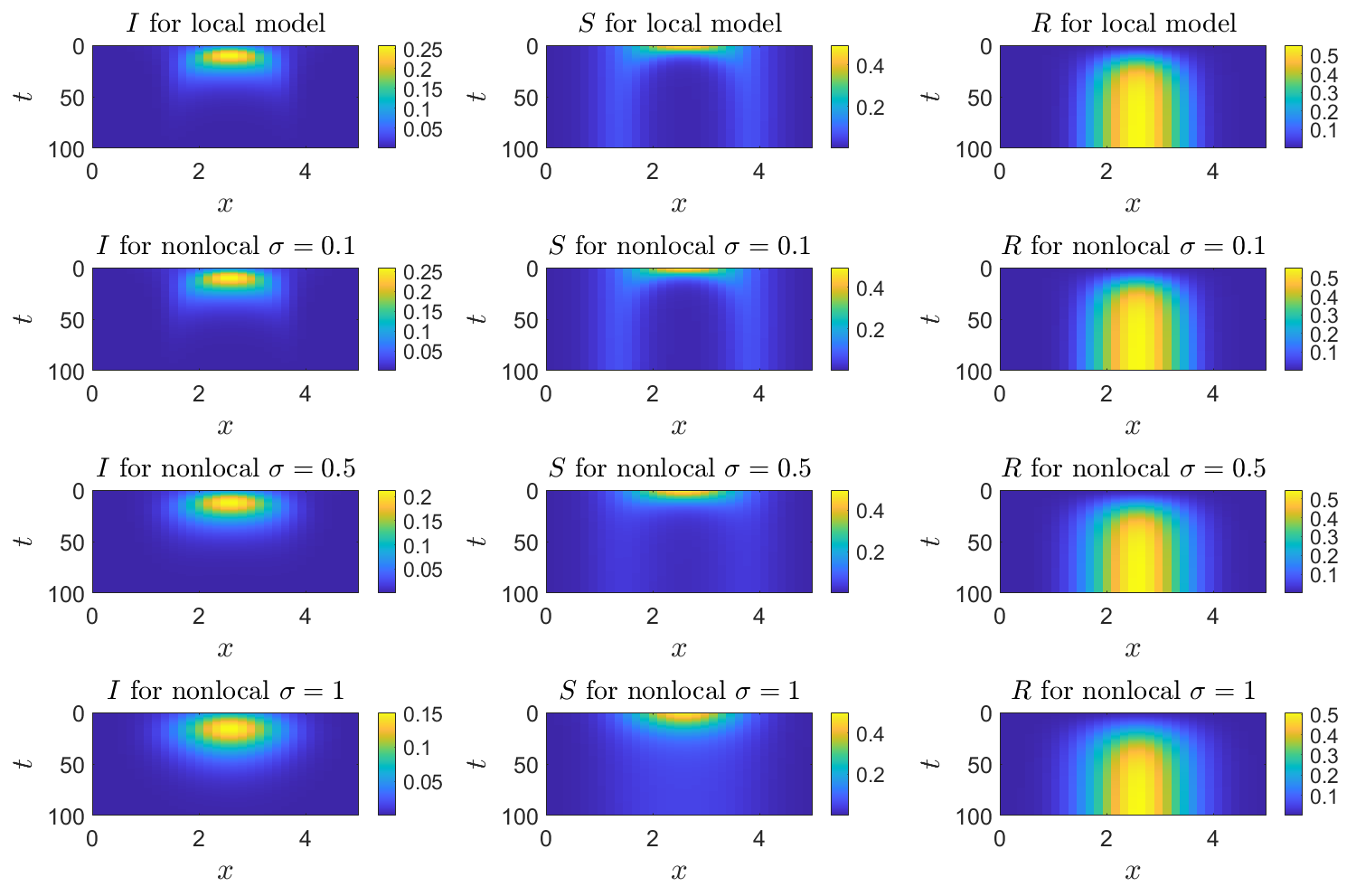}
    \caption{Evolution dynamics of the 1d SIR models with co-localized Gaussian initial conditions \eqref{eq: colocalization}. Across all simulations, we fix $\beta = 0.8$, $\gamma = 0.1$, and $\mu = \frac{1}{2}\beta\sigma^{2}$. We use $\sigma = 0.1$ for the local model (top panel) and use $\sigma = 0.1$, $0.5$, and $1$ for the nonlocal models (bottom three panels).
    }
    \label{fig: 1d-SIR-models-colocalized}
\end{figure}

We show the evolution dynamics of the spatial local and nonlocal SIR models in Figure \ref{fig: 1d-SIR-models-colocalized}. For the local model and the nonlocal model with $\sigma=0.1$, we notice waves of infection moving away from the mass center at $x=2.5$. However, for the nonlocal models with $\sigma=0.5$ and $\sigma=1$, we observe a diffusion of infection. 
Additionally, in the local model, the susceptible population shows a low-density region near $x=2.5$ for $t>20$. This occurs because, in the local model, infected individuals are more likely to infect their nearby individuals, leading to significant infections and a reduction in population near $x=2.5$. On the other hand, the nonlocal model with $\sigma=1$ exhibits a more homogeneous population distribution after $t>20$, meaning that some susceptible individuals around $x=2.5$ remain uninfected until the pandemic ends. Infected individuals are less infectious to their neighbors compared to the local models, so some susceptible individuals never get infected before the nearby infected individuals near $x=2.5$ recover.

\subsection{A 2d example with Gaussian initial densities}
\begin{figure}[htp]
    \centering
    \includegraphics[width=0.22\linewidth]{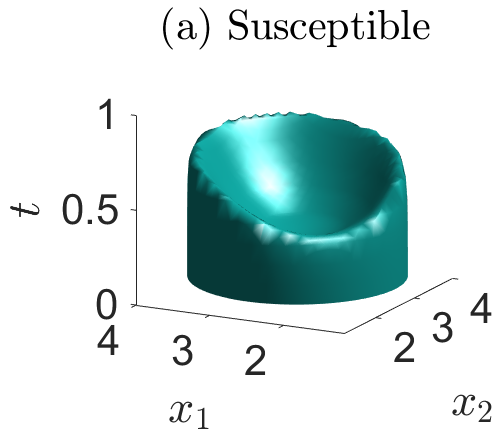} \qquad 
    \includegraphics[width=0.22\linewidth]{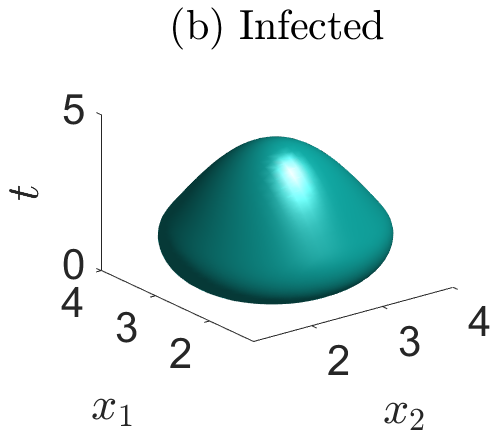} \qquad
    \includegraphics[width=0.22\linewidth]{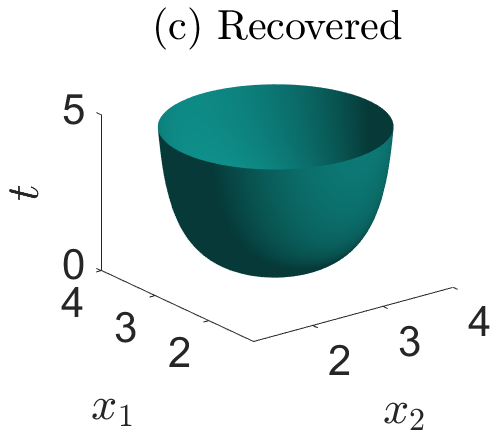}
    \caption{The iso-surface representation of the evolution dynamics of the 2d weakly-nonlocal SIR model \eqref{eq: nonlocal_SIR} with the width parameter $\sigma = 0.1$. The infection rate $\beta = 100$, the recovery rate $\gamma = 0.5$, and the initial infection ratio $\eta = 0.01$. The three population densities ($S(x,t)$, $I(x,t)$, and $R(x,t)$) have constant function values $0.054, 0.038$, and $0.045$ on their respective iso-surfaces.
    }
    \label{fig:12}
\end{figure}

We investigate a 2d weakly nonlocal SIR model with $\sigma=0.1$ and plot the iso-surfaces of three densities in Figure \ref{fig:12}. The initial densities for the susceptible and infected populations are: 
\begin{equation}\label{eq: 2d Gaussian density}
    \begin{aligned}
    S_0(x_1,x_2) &\propto \exp\left[-\left(x_1-2.5\right)^{2} - \left(x_2- 2.5\right)^{2}\right], \quad 
    I_0(x_1,x_2) \propto \exp\left[-\left(x_1-1\right)^{2} - \left(x_2- 1\right)^{2}\right].
    \end{aligned}
\end{equation} 
The susceptible population has a Gaussian initial density. As shown in Figure \ref{fig:12}(a), the susceptible population decreases over time, with the center decreasing faster than the perimeters. In the weakly nonlocal models, infected individuals are more infectious to their neighbors compared to the nonlocal models. This leads to a rapid spread of infection in high-density regions, causing a faster decay in the center population. 
Additionally, since infections are initially located near $(1,1)$, the susceptible population near $(1,1)$ gets infected sooner compared to regions farther away, forming an asymmetric pattern in the susceptible iso-surface. The infected population initially increases, but around $t=1$, the recovery process begins to dominate, leading to a decrease in the infected population.

\end{appendices}

\bibliographystyle{abbrv}
\bibliography{references}

\end{document}